# A nano polymer aggregate on a substrate: A Theoretical study

**Pramod Kumar Mishra**

Department of Physics, DSB Campus, Kumaun University, Nainital,Uttarakhand (INDIA)

**Abstract**. We use lattice model to investigate the thermodynamics of an ideal semi-flexible polymer chain at the nano scale. The polymer chain is assumed to polymerize on the nano size substrate and the conformations of the chain were realized on the substrate using a square lattice. Our analytical estimates show clear distinctions in the thermodynamics of the polymer chain from its corresponding bulk . The persistent length versus stiffness curve varies linearly at the possible small length scale but in the bulk this variation is gradual. The polymer chain is confined from each direction and therefore, the nature of the force of confinement on a flexible and a semi-flexible chains has also been discussed for the proposed nano scale confinement.

## 1. Introduction

The many body systems show different characteristics at the nano-length scale than its corresponding bulk state [1]; it is understood also that the materials possess useful and distinct properties at the nano regime, therefore there is much interest among the researchers to investigate the properties of the materials at the nano length scales using experimental as well as theoretical methods [1-2]. There are a variety of useful applications of the nano materials, and therefore, there is reasonable funding for research projects in the area of the nano materials. The nano-materials are specially synthesized to show more surface effects and it is due to a fact that the number of molecules/atoms at the surface increased to several folds from the bulk; and thus presence of more surface atoms/molecules are responsible for increased surface effects and this reason leads distinct properties of the materials at the nano scales. Thus, at the nano scales the properties of materials; for example the thermal, mechanical, optical properties may change to many extents from corresponding bulk state of the materials [1-5].

There are limited theoretical studies found in the literature regarding investigations on the behavior of the polymeric materials at the nano length scales [6]. Therefore, we consider the case of an infinitely long semi-flexible Gaussian polymer chain and this chain is assumed to make of identical repeat units (monomers). The Gaussian semi-flexible homo-polymer chain is confined on a square shapes substrate of *NXN* monomers size, where (N=1,2,3..5); and we derive the thermodynamics of an infinitely long ideal (the Gaussian) semi-flexible polymer chain using the methods of recursion relations [7]. The macromolecules (the Gaussian chain) is thus confined in a fairly small area, and it is due to the excellent elastic properties of such macromolecules. Therefore, we have chosen an infinitely long ideal polymer chain to study the thermodynamics of such chain under proposed geometry.

The manuscript is organized as follows: we describe the model of an ideal polymer chain using square lattice in section two, and the method of calculations is also described in section two in the brief for the sake of completeness. The results are discussed in section three; and also in section three, we report the critical value of the monomer fugacity for the polymerization of an infinitely long semi-flexible polymer chain. We calculated the persistent length and the elastic force of the confinement in the proposed geometry of the semi-flexible chain; while the summary and the conclusions were highlighted in the section four.

## 2. Model and method

A lattice model of an ideal polymer chain is extensively used to describe the thermodynamics of a semi-flexible homo-polymer chain in the bulk [7-9]. However, thermodynamics of an ideal chain in the nano-area is not well understood yet. Therefore, we consider an ideal chain on a two-dimensional substrate and the area of the substrate is in the nano length scale. A square lattice is used to mimic the conformations of the polymer chain in the nano-regime. The conformations of the polymer chain is obtained using the random walk and the walker is allowed to take steps along ±x and ±y directions. There are two bending angles ($\theta$) permitted to the walker while enumerating the conformations of the chain; i. e. *90º* and *180º*; while the bending energy is the function of the bending angle as $E_b(\theta) = E(1 - \cos\theta)$. Thus, we have Boltzmann weights $k$ (=$e^{-\beta E}$) and $k^2$ corresponding to the bending angles *90º* and *180º*, respectively.

The Bio-molecules are the semi-flexible polymer chain and they have excellent elastic properties so that such macromolecules can be easily confined to the nano length scales [10-12]. Therefore, we consider a semi-flexible polymer chain in the thermodynamic limit to study the role of the nano length scale confinement on the thermodynamics of an ideal semi-flexible polymer chain. For example, the basis vector of the square lattice is *5Å*, say; the substrate is square shaped and thus we may be able to study theoretically the behavior of the semi-flexible polymer in the area ~*625X10$^{-20}$m$^2$*. Such an analytical study may be useful to investigate an interesting and distinct insights into the thermodynamics of the macromolecules at the nano-length scales. There is seventeen monomers long walk of the confined semi-flexible ideal polymer chain which is shown schematically in figure no. 1. The walk of the semi-flexible chain is grafted at a point *O*, and the symbol $X_i$ (where *i* is an integer, and it is related to position of the walker with respect to the boundary of the nano substrate and also to the direction of stepping of the walker) corresponds to the sum of the Boltzmann weight of all the walks with its first step along the x-direction. However, a detailed mathematical structure of the this report is planned to be published elsewhere. The grand canonical partition function [$G(g,k)$] of the confined semi-flexible polymer chain may be written in general as,

$$G(g,k) = \sum_{All\ walks\ of\ N\ monomers} \sum_{N=1}^{\infty} \sum_{B=0}^{\infty} g^N k^B \qquad (1)$$

Where *g* is the fugacity corresponding to each monomer and *k* is the Boltzmann weight corresponding to bending energy of the semi-flexible chain. The number of monomers in the chain is *N,* while number bends in the chain is *B;* (where one bend corresponds to *90º* and *180º* bending angle corresponds to the number of bends equal to 2, and it has been done so merely to ensure distinction in the possible bending angles for the Gaussian chain) and *B* may be greater than the length (*N*-monomers) of the polymer chain.

## 3. The results

The components of the grand canonical partition function are written as [7-9,13-16],

$$X_1 = g + gX_2 + k^2 gX_2 + gkX_3 \qquad (2)$$

$$X_2 = g + gkX_1 + gk^2 X_1 \qquad (3)$$

$$X_3 = g + (g + 2gk + k^2 g)X_4 \qquad (4)$$

$$X_4 = g + 2gkX_2 + gk^2 X_3 \qquad (5)$$

We can solve the above four equations (*i. e.* Eqn. nos. 2-5) to obtain the partition function of an ideal semi-flexible polymer chain for the case when the polymer chain is polymerized in the nano area of size 2X2 unit.

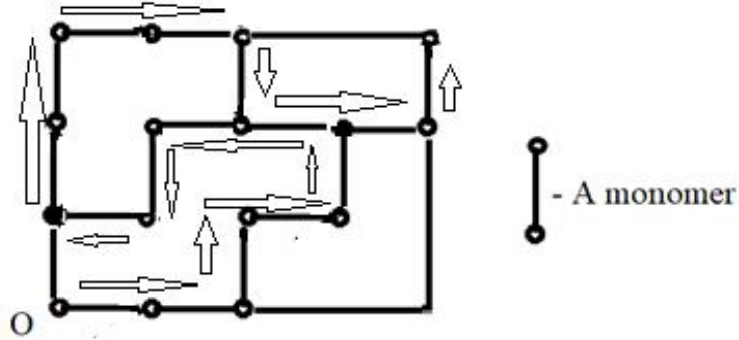

Figure No. 1: A seventeen monomers long semi-flexible ideal chain is shown schematically in this figure. There are eleven bends in the chain, and a monomer is also shown schematically in this figure. The substrate has nano area and the area is equal to 4X3 units. The length is measured in the units of the monomer size.

$$G^{2X2}(g,k) = \frac{(2g(1-g^2(-1+k)k(1+k)^2-g^3(-1+k)k^2(1+k)^3+g(1+k+k^2)))}{(1+g^4(-1+k)k^3(1+k)^4-g^2k(1+2k+3k^2+2k^3))} \qquad (6)$$

We determine a condition for the polymerization of an infinitely long ideal chain in the nano region of size *2X2* unit, and the critical value of the monomer fugacity ($g_c$) is shown in figure no. (2) and its numerical values are shown in table no. 1 for the different areas on the nano scales, and also for the bulk case. The variations of the critical values of the step fugacity are shown in figure no. 2, and this graph clearly shows a distinction from the bulk value of the monomer fugacity on the polymerization of an ideal polymer chain in the thermodynamic limit.

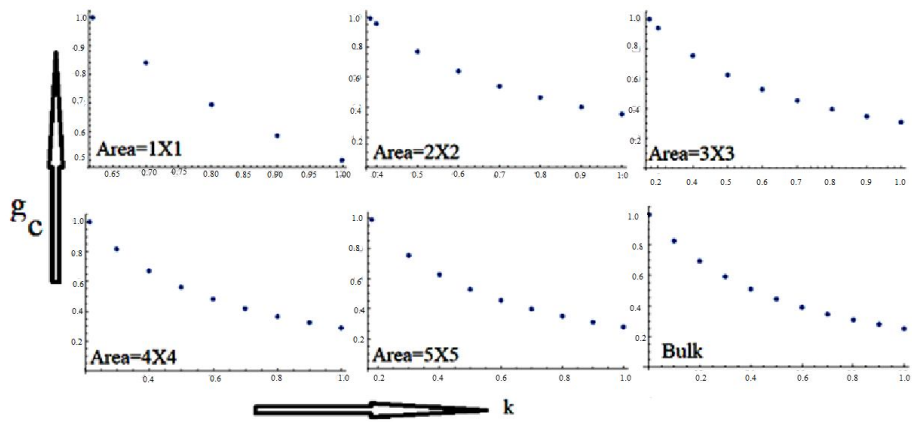

Figure No. 2: This figure shows the critical value of the monomer fugacity on the polymerization of an infinitely long semi-flexible chain on the nano-size substrate. The Nano size varies from one to five monomer size and chosen substrate size is square in the shape. The critical value of monomer fugacity

for bulk (N→∞) was also shown for the sake of comparison. It is clear from this figure that the stiff chain cannot polymerize in the nano area.

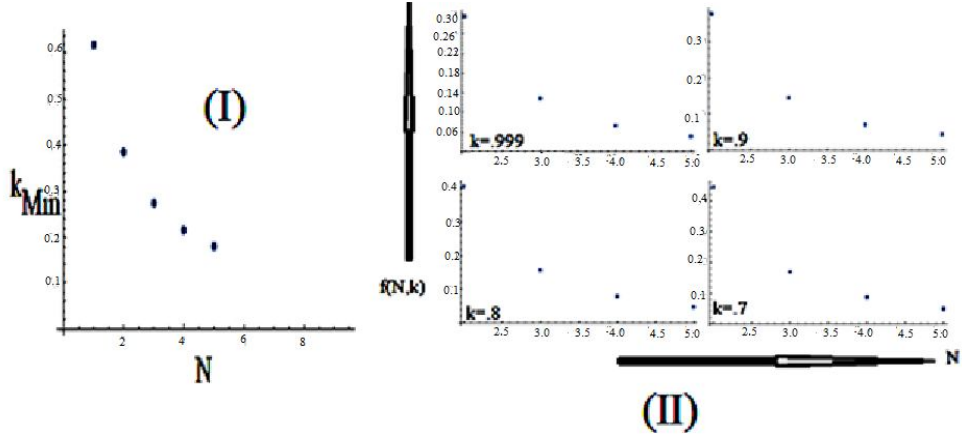

Figure No. 3: We have shown the minimum value of stiffness weight ($k_{Min}$) of an infinitely long ideal chin in the 3(I) for the values above $k_{Min}$ a polymer chain is polymerized on the nano-substrate. The force of confinement acting on each monomer of an infinitely long ideal chain is also shown in 3(II) for the flexible ($k=1$) and the semi-flexible polymer chains ($k=.9, .8$ & $.7$).

It is seen that the stiff chain polymerization is not possible on the nano substrate [$k \geq k_{Min}(g=1)$] and a possible minimum value of the stiffness [$k_{Min}(g=1)$] weight is shown in figure no. 3(I) for the nano-area confinement.

The force of confinement is also shown in figure no. 3(II) for a set of the values of the stiffness of the semi-flexible polymer chain and the flexible polymer chain ($k=1$). The force of the confinement [17] is calculated using the following equation,

$$f(N,k) = \frac{Log[(N-1,k) - Log(N,k)]}{1} \qquad (7)$$

Where, N=2,3,4 and 5.

We have plotted the value of the persistent length [13-16] of the nano-polymer aggregate in figure no. 4. It is shown that the persistent length of the chain is smaller than the length of confinement of the polymer chain.

## 4. The summary and the conclusions

We consider polymerization of an ideal polymer [7-9] on a nano substrate, and we realize the conformations of the Gaussian semi-flexible polymer chain on a square lattice. A method of the recursion relation [7, 13-16] is used to obtain the grand canonical partition function of an ideal polymer chain for the proposed nano scale confinement. The singularity of the partition function [$G(g,k)$] is obtained to derive the thermodynamics of the chain for its bulk and the confined cases; Where the nano-substrate has dimension *NXN* (where N=1,2,...5 monomers, and the size of the monomer and the basis vector of the square lattice is same) monomers for the proposed study.

The singularity of the partition function gives us the critical value of the monomer fugacity ($g_c$) for the polymerization of an infinitely long semi-flexible polymer chain on the nano substrate, and we have also reported the value of ($g_c$) for the bulk case. The variation of the critical values of the monomer fugacity is shown in figure no. (2), and these values of the monomer fugacity ($g_c$) are shown in table no. (1) to discuss the effect of such confinement of an ideal semi-flexible chain at the nano-scale confinement. It has been found that the stiff chains are not polymerized on the nano substrate.

The monomer fugacity has a distinction from its bulk value and it is clearly seen from figure no. (2). The possible minimum value of the chain stiffness [$k_{Min}(g_c=1)$] is also shown in figure no. 3 (I) to understand the effect of such confinement. The force of confinement acting on each monomer of the Gaussian chain is also shown in figure no. 3 (II) and in the present report chain is confined to nano length scale from all the possible directions.

**Table No. 1:** This table shows the critical value of the monomer fugacity regarding the polymerization of an infinitely long semi-flexible homo-polymer chain. The critical value $g_c$(bulk) corresponds to the infinite size of the substrate along both the directions. While $g_c(NXN)$ [where $N=1,2,3,4$ & $5$] corresponds to the critical value of the monomer fugacity corresponding to the substrate size varying from one to five monomers.

| K | $g_c$(bulk) | $g_c$(1X1) | $g_c$(2X2) | $g_c$(3X3) | $g_c$(4X4) | $g_c$(5X5) |
|---|---|---|---|---|---|---|
| 0.001 | 0.998 | - | - | - | - | - |
| .1 | 0.826 | - | - | - | - | 0.992 ($k=.18$) |
| .2 | 0.694 | - | - | .999($k=.275$) | 0.999($k=.2156$) | 0.943 |
| .3 | 0.592 | - | 0.989 | 0.94 | 0.819 | 0.754 |
| .4 | 0.51 | - | 0.955 | 0.756 | 0.671 | 0.625 |
| .5 | 0.44 | - | 0.77 | 0.626 | 0.563 | 0.529 |
| .6 | 0.39 | 0.998($k=.619$) | 0.638 | 0.53 | 0.482 | 0.455 |
| .7 | 0.346 | 0.840 | 0.539 | 0.455 | 0.417 | 0.397 |
| .8 | 0.309 | 0.694 | 0.463 | 0.396 | 0.366 | 0.349 |
| .9 | 0.277 | 0.585 | 0.402 | 0.348 | 0.324 | 0.310 |
| .999 | 0.25 | 0.500 | 0.354 | 0.309 | 0.289 | 0.278 |

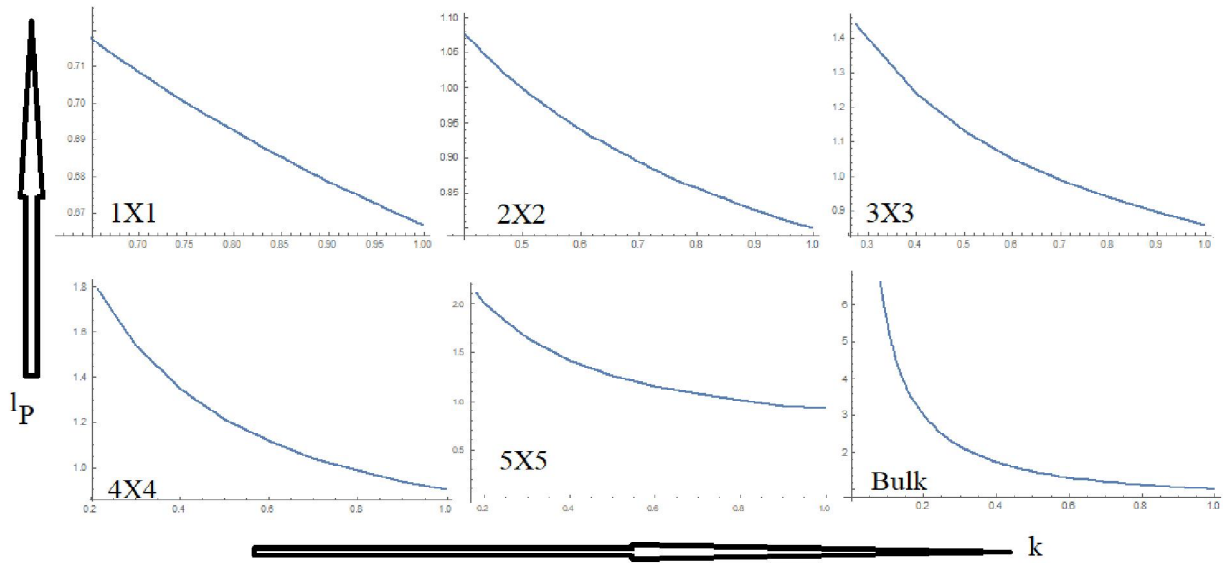

Figure No. 4: We show the value of the persistent length of the semi-flexible polymer chain for the nano size confinement. It is shown that an average length between two successive bends ($l_P$) of the chain is less than the confinement length ($N$), when the chain is polymerized in the nano area $NXN$. We have also shown the variation of the persistent length of a semi-flexible polymer chain as the function of the stiffness ($k$) of the chain. The results for chosen values of $N$, as well as for the case of the bulk are also shown in this figure.

The persistent length [13-16] of the Gaussian chain semi-flexible is shown in figure no. 4 for the cases of the nano scale confinement along with the value of the persistent length for the bulk. It is seen that the persistent length is smaller than the length of the confinement of the Gaussian semi-flexible chain. The persistent length has a distinct nature of its variation with the chain stiffness than the bulk value of the persistent length of the chain.

Thus, the Gaussian semi-flexible polymer chain is polymerized on the nano size substrate. However, stiff Gaussian chains [k<$k_{Min}(g_c=1)$] cannot polymerize on the nano size substrate. There is large entropic force acting on each monomer of a flexible polymer chain than a stiff chain. The entropic force decreases as the length of the confinement increases. The persistent length of the polymer chain is smaller than the length of confinement of the chain. The length of the confinement of the chain is nothing but the maximum size of the substrate along any of the *x* or *y* direction. The method of the present calculations can be easily extended to understand the behavior of a self-avoiding polymer chain to discuss the thermodynamics of the polymer chain at the nano-length scale, provided at least one direction of the polymer chain may have an infinite extension. The method of present calculations can also be use to derive thermodynamics of a short-chain (i. e. for the Gaussian as well as the self-avoiding polymer chains) for the nano scale confinement of the macromolecule.

**References**
[1]   Jeremy Ramsden 2016 *Nanotechnology: An introduction*, (Elsevier Science, 2nd Edition).
[2]   Amretahis Sengupta and Chandan Kumar Sarkar ed 2015 *Introduction to nano: Basics to Nanoscience and Nanotechnology*.
[3]   Allhoff Fritz, Lin Patrick and Moore Daniel 2010 *What is nanotechnology and why does it matter?: From science to ethics* (John Wiley and Sons) 3–5.
[4]   Prasad S K 2008 *Modern Concepts in Nanotechnology*, (Discovery Publishing House) pp 31–32.
      K Jennifer 2006 (June) *"Nanotechnology" National Geographic 98–119.*
[5]   Polymer-*based nanocomposites for Energy and Environmental Applications* 2018 M. Jawaid and M. M. Khan ed (Woodhead Publishing, Elsevier Ltd).
[6]   Alfred J Crosby and Jong-Young Lee 2007 Polymer Reviews **47(2)** 217-229.
[7]   Privman V, Svrakic N M 1989 Directed models of polymers, interfaces, and clusters: scaling and finite-size properties, (Berlin Springer).
[8]   Vanderzande C, Peter G, Yeomans J ed 1998 Lattice models of polymers Cambridge University Press.
[9]   D´e Bell K and Lookman T 1993 Rev. Mod. Phys., **65** 87.
[10]  Hansma H G, *etal* 1997 Structural Biology, **119** 99-108.
[11]  Kishino A and Yanagida T 1998 Nature (London) **34** 74.
[12]  G V Shivashankar and A Libchaber 1997 Appl. Phys. Lett. **71** 3727.
[13]  P K Mishra, Kumar S and Singh Y 2003 Phys A **323** 453-465.
[14]  P K Mishra 2010 J Phys: Cond Matt. **22** 155103.
[15]  P K Mishra 2014 Condens Matter Phys. **17(2)** 23001.
[16]  P K Mishra 2015 Phase Transitions **88(6)** 593-604.
[17]  P K Mishra 2017 Ind J Phys **91** 1297-1304.